\documentclass[twocolumn,noshowpacs,preprintnumbers,amsmath,amssymb]{revtex4}

\usepackage{verbatim}

\usepackage{graphicx}
\usepackage{dcolumn}
\usepackage{bm}
\usepackage{wrapfig}

\begin{document}

\title{Synthesis of InP nanoneedles and their use as Schottky devices}

\author{Tim Strupeit}
\author{Christian Klinke}
\author{Andreas Kornowski}
\author{Horst Weller}
\email{weller@chemie.uni-hamburg.de}
\affiliation{Institute of Physical Chemistry, University of Hamburg, Germany}

\begin{abstract} 

Indium phosphide (InP) nanostructures have been synthesized by means of colloidal chemistry. Under appropriate conditions needle-shaped nanostructures composed of an In head and an InP tail with lengths up to several micrometers could be generated in a one-pot synthesis. The growth is interpreted in terms of simultaneous decomposition of the In precursor and in situ generation of In and InP nanostructures. Owing to their specific design such In/InP nanoneedles suit the use as ready-made Schottky transistors. Their transfer and output characteristics are presented.

\end{abstract}

\maketitle
Wet-chemically prepared semiconducting nanostructures are promising candidates for future electronic devices, not in the least due to their ability to accommodate whole device structures in a single nano-object~\cite{1}. Additionally the electronic properties of those nanostructures can be modified by size and dimensionality owing to quantum confinement effects as has been demonstrated for zero- and one-dimensional CdSe quantum structures by optical spectroscopy~\cite{2,3,4,5,6,7,8}. The one-dimensionality of nanorods affords new properties like the polarized emission under photoexitation and electroluminescence~\cite{9,10}. A major drawback of the most frequently investigated materials is their chemical composition (chalocogenites of cadmium, lead, or mercury) and the toxicity resulting thereof. Presently, the most attractive and the less toxic alternative materials is InP, and, consequently, the corresponding nanowires are considered as building blocks for novel types of nanowire-based photo devices and solar cells~\cite{11}. On the other hand, the synthetic protocols for high quality InP nanostructures are still less developed than for II-VI and IV-VI materials, and also
shape control is best established in II-VI systems (e.g. CdSe). Here the growth is controlled by the amount and type of added ligand molecules and allows the synthesis of rods, dots, or tetrapods~\cite{4,9,12}. The underlying growth mechanism is limited to semiconductors with an anisotropic crystal structure like wurzite since they expose facets with chemically distinguished reactivity; that is, the ligands preferentially bind to one family of surface planes resulting in a preferential growth in one direction. III-V semiconductors, however, possess a cubic zincblende lattice
structure; that is, the required anisotropy of chemically different surfaces is not given. In this case another growth mechanism, the solution-liquid-solid (SLS) mechanism, can be exploited~\cite{13}. A liquid metal droplet acts as seed and catalyst for the crystal growth. This
mechanism was found for a few systems establishing III-V semiconductor rods and wires. Commonly used seeds are gold, bismuth, silver, or indium~\cite{4,7,14,15}. Usually such methods represent a two-step synthesis. During the first step the metal seeds are produced by decomposing an organometallic precursor, whereas the dehalosylation reaction of InCl3 or InAc3 with tris-(trimethylsilyl)-phosphine (P(Si(CH)3)3, TMSP) in a highboiling coordinating solvent (trioctyl phosphine (TOP) or trioctylphosphine oxide (TOPO)) occurs during the second step. This reaction enables the synthesis of nanowires and nanorods of some micrometers in length~\cite{16}. Since the diameter of the growing nanowires was found to depend on the size
of the primarily formed metal seeds, recent investigations focus on the diameter control of these seeds. 

Here, we present a new, rapid, versatile, and less toxic reaction route for the synthesis of needle-shaped InP nanowires with lengths of several micrometers. We also show that this procedure is capable of forming a readily assembled Schottky diode and transistor device and present corresponding
current-voltage measurements on such individual structures.

\begin{figure}[htbp]
  \centering
  \includegraphics[width=0.45\textwidth]{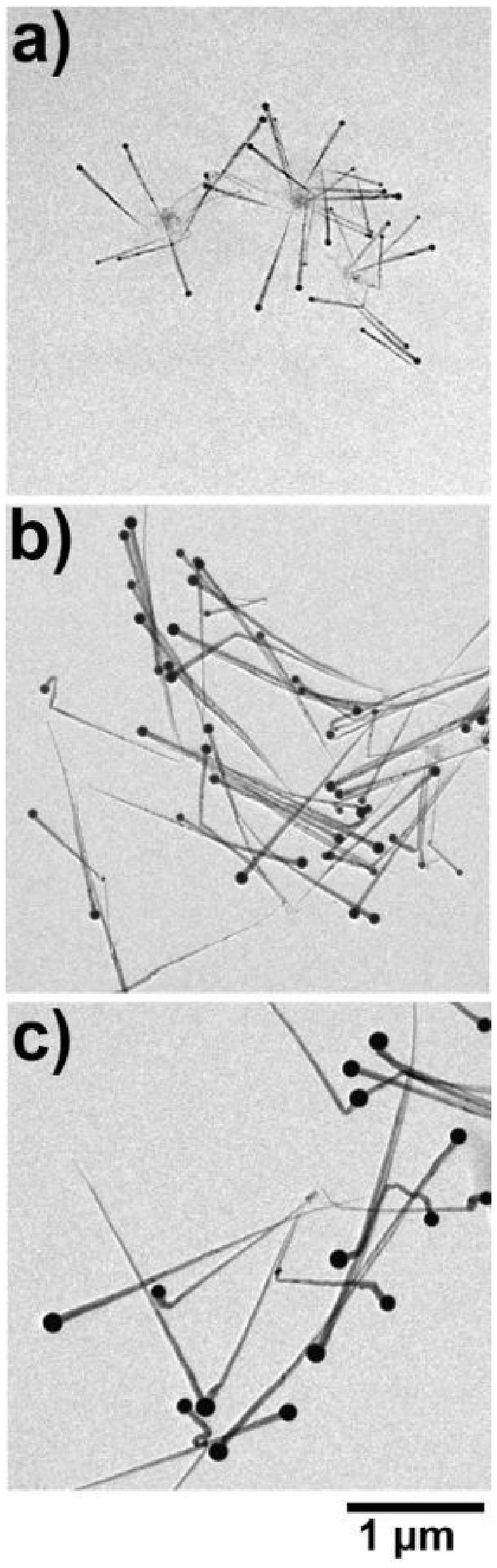}
  \caption{TEM overview of In/InP needles synthesized at 200$^{\circ{}}$C. Samples taken after (a) 18, (b) 21, and (c) 31 min.}
  \label{Fig1}
\end{figure}

In contrast to the well established dehalosylation reaction, where the harmful and highly reactive phosphorus component is applied, we use here the low reactive TOP and the moderately reactive trimethyl-indium precursor as already described in references~\cite{17,18,19} for the synthesis of spherical InP nanoparticles. This reversal in reactivity has dramatic consequences on nucleation and growth during the SLS process. In(CH3)3 decomposes already at temperatures above approximately 125$^{\circ{}}$C, whereas pure TOP is stable even above 300$^{\circ{}}$C. Thus, the first step is the formation of indium nanodroplets (mp of bulk In, 156$^{\circ{}}$C), which in turn act as catalyst for TOP decomposition. Decomposition of TOP produces surface-bound P species, which can in turn react with In to form InP. Once an InP nucleation seed is formed at the surface of a droplet, it is fed additional indium and phosphorus via this same surface process; that is, the seeded growth starts. To obtain just one uniform InP wire out of each In droplet, TOP decomposition and the flux of phosphorus and indium or InP monomers to the growth seed must be balanced. Since the activation energies for the involved processes are expected to be different, temperature should play a crucial role for the morphology of the final product. At temperatures below the optimum T$_{opt}$, we expect TOP decomposition and phosphorus flux as very slow favoring random nucleation and growth and thus a polymorph final product, whereas at T $>$ T$_{opt}$ nucleation of InP seeds should be favored leading to the formation of several wires out of one In nanodroplet or under conditions of high reactivity and high phosphorus flux to the dissolution of the droplets and formation of InP nanoparticles. As outlined in detail in the Supporting Information the expected temperature dependence could be verified, whereby an optimum temperature around 200$^{\circ{}}$C was found.

\begin{figure}[htbp]
  \centering
  \includegraphics[width=0.45\textwidth]{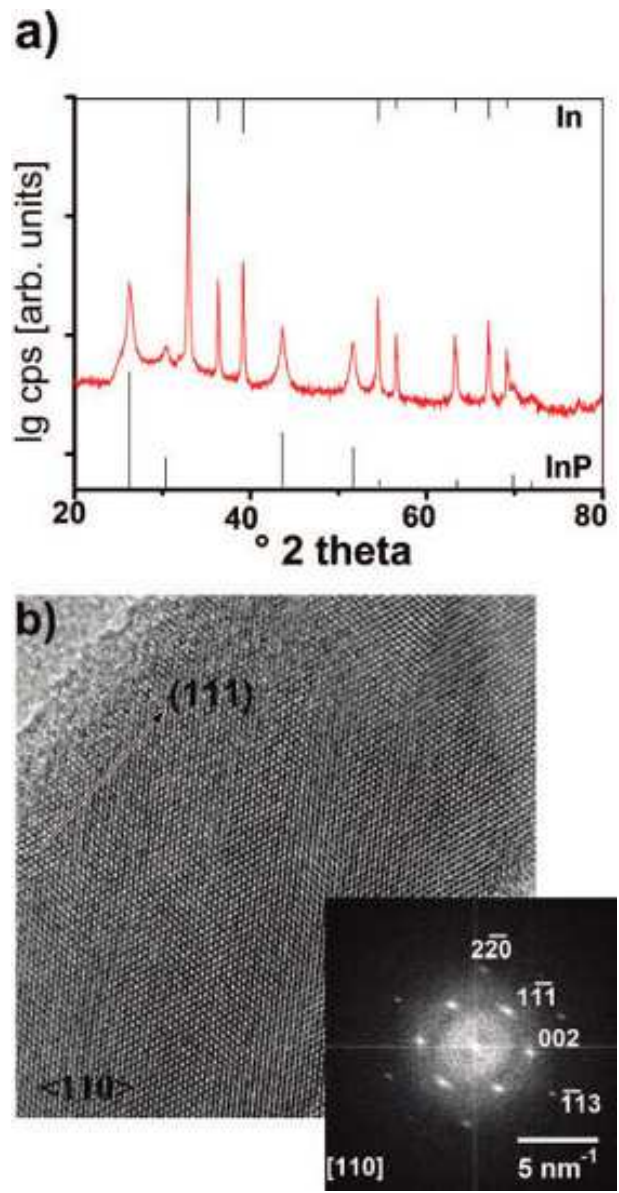}
  \caption{(a) XRD pattern of a In/InP needle sample (red), pure In (top), and pure InP (bottom); (b) HRTEM image of a InP needle in $<$110$>$ orientation with a growth direction along the (111) axis (arrow) and the FFT (inset) with the corresponding indexing.}
\end{figure}

We further investigated the wire growth at 200$^{\circ{}}$C after various reaction times. During the first quarter of an hour no significant color change was observed and no particles could be isolated. After 18 min, however, a fast color change of the solution occurred. At this time small nanoneedles with a maximum length of 700 nm could be separated (Figure 1a). The diameter of the In seeds was below 100 nm. Within 3 min the needles
grew to some micrometers in length and also the seed diameter increased (Figure 1b). The diameter of the InP wire was always smaller than the diameter of the In seed (ratio approximately 1:3). After 21 min the needles were grown to a length of several micrometers. Further increase of reaction time
to 31 min mainly increased the size of the In spheres and resulted in a more polydisperse sample, in which often more than one InP wire was attached to an In sphere (Figure 1c). This might be understood by a fusion of the molten In droplets of already formed nanoneedles. 

Astonishingly, it was not possible to isolate individual indium seeds without attached InP. We assume that the catalytic cleavage of the phosphorus-carbon bond of TOP starts immediately after the In seed formation. The freshly produced free phosphorus instantly reacts to InP by the SLS mechanism. Because of the high excess of TOP in the reaction mixture, the surface of the formed In spheres is covered with TOP which, on one hand, stabilizes the In particles and acts, on the other, as the phosphorus source for needle growth. In a blank experiment, in which TOP was replaced by ODE, only bulk In was formed. 

The In/InP nanoneedles were characterized by X-ray diffraction (XRD). Figure 2a shows the XRD pattern of nanoneedles synthesized at 300$^{\circ{}}$C after 2 min of reaction time. The reference pattern of InP (PDF 73-1983) and In (PDF 85-1409) are also depicted in the figure. Sharp In reflexes (corresponding to the tetragonal lattice of In) accompanied by smaller and broader InP reflexes (zinc blende) are clearly visible. Energy-dispersive X-ray spectroscopy (EDS) of the needles and the seeds show that the seeds consist mainly of In, whereas the nanoneedles are composed of InP with an In to P ratio of about 60:40 indicating an In-rich surface. High-resolution transmission electron microscopy images (HRTEM, Figure 2b) prove the high degree of crystallinity but exhibit some stacking faults within the InP wire. Also visible is an amorphous surface layer, which might be responsible for the nonstoichiometric In to P ratio. The main growth direction of the needles is along the (111) axis. Figure 2b displays the InP needle in the $<$110$>$ projection. The corresponding fast Fourier transformation (FFT) (inset) exhibits the expected distances for the
[111], [200], [220], and [311] lattice planes. Furthermore, we identified the crystal planes at the interface between the InP tail and the In seed. Whereas in general the In spheres were polycrystalline, we found a favored orientation of the [002] planes of In at the interface between the In sphere and the InP wire. Since indium is liquid at the reaction temperatures and crystallizes only in a post preperative step, this orientation is probably directed by the [111] interface planes of InP. We also found distances of about 2.93 A at the indium surface and in the vicinity of the InP wire. This distance matches no In spacing but can be assigned to the [200] spacing of InP. This observation supports the assumption that wire growth is fed via a catalytically formed thin surface layer of InP.

\begin{figure}[htbp]
  \centering
  \includegraphics[width=0.45\textwidth]{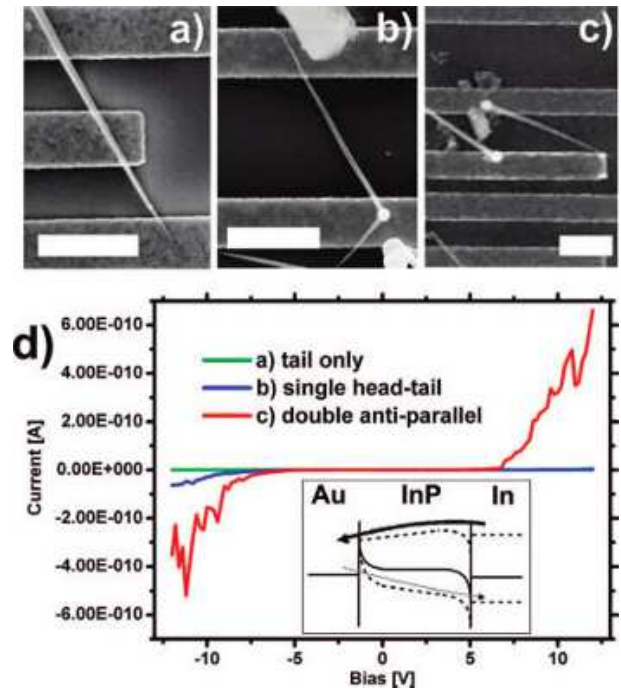}
  \caption{(a) SEM mircographs of three different configurations of In/InP nanoneedles attached to gold leads: (a) only the InP tail bridges the
leads, (b) a single nanoneedle attached with In head and InP tail, and (c) two needles attached in a antiparallel fashion (scale bar corresponds
to 1~$\mu$m). The corresponding transport characteristics are shown in panel d. The inset shows the band alignment for the classical (thick arrow) and the tunnelling Schottky barrier.}
\end{figure}

Because of their head-tail morphology the nanostructures shown in Figure 1b are predestined to be used as electronic building blocks. A view to the involved energy levels, workfunctions, and electron affinities reveals that the In/ InP interface of the nanoneedles forms an Ohmic contact. On the other hand a possible connection of the InP tail to a gold electrode would render a Schottky barrier. The electronic structure of such a device is depicted in the inset in Figure 3d (solid line). Applying a bias at the In contact side (gold contact is grounded) results in an enhanced (upper dashed line) or diminished effect of the Schottky barrier (lower dashed line). In the first case current can flow through the device when the bias voltage is
large enough to drive electrons over the Schottky barrier. In the latter case, a reversal of the bias voltage results in flow of current through the composite structure due to tunneling through the Schottky barrier at the Au/InP contact. Considering this configuration a nanoneedle bridging two gold leads would thus result in a ready Schottky diode structure. The three most interesting possibilities to attach In/InP nanoneedles to gold leads are shown in the SEM micrographs in the upper part of Figure 3 (which were taken after the electrical measurements). The dropcasted nanoneedles can either bridge two gold contacts (predefined by e-beam lithography on SiO2) only with their tail InP part (a), with the In head on one side and the InP tail on the other (b), or more than one nanoneedle can bridge the leads. Here we show the antiparallel configuration with two nanoneedles (c). In the first case, having two Schottky contacts (InP/Au), one at each contact, we did not measure any significant current varying the bias between -12 and +12 V. The signal remains in the noise range (a log-scale plot can be found in the Supporting Information, Figure S3). In the second, prototypical case, we obtained an asymmetric I-V curve as expected for such Schottky diode devices as described above. At negative voltages starting from -3 V the current reaches 62.9 pA at -12 V. This is due to the overcoming of the Schottky barrier (classical Schottky diode). At positive voltages, however, the obtained current remained more than an order of magnitude lower. This current is probably due to electron tunneling through the Schottky barrier (tunneling Schottky diode). The asymmetric behavior is understood in terms of the asymmetric electronic band alignment at the contacts, whereas an almost symmetric I-V characteristic can be observed with a two-nanoneedle configuration with antiparallel orientation. In each direction always one element blocks the current and the other one is in (classical) forward-bias mode.

\begin{figure}[htbp]
  \centering
  \includegraphics[width=0.45\textwidth]{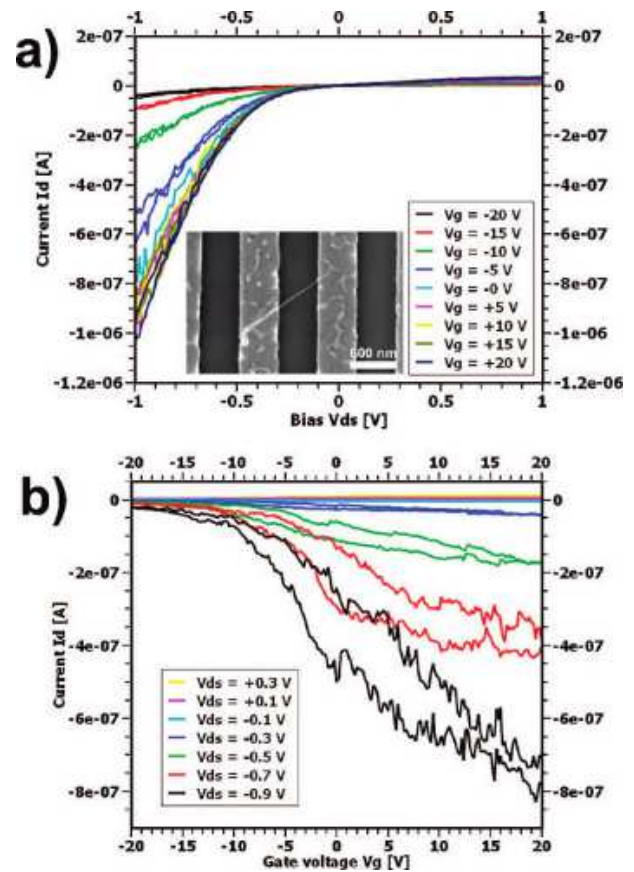}
  \caption{(a) Output characteristic of the nanoneedle in a three-terminal field-effect transistor configuration. Inset: SEM micrographs of the device. (b) Transfer characteristic at different bias voltages (bidirectional sweeps).}
\end{figure}

In the cases discussed above, the contact resistances required relatively high bias voltages. To improve this situation we spread the nanoneedles first on the Si/SiO2 substrate followed by subsequent deposition of the gold electrodes by e-beam lithography. Because of the thermal energy of the deposited gold, this process produces greatly improved contacts between the gold electrodes and nanostructures. The output characteristic (Figure 4a) shows the typical diode behavior of the device with practically no hysteresis. At much lower negative biases than before, the device switches on and at positive biases it blocks the current. The Si substrate can be used as a gate electrode to electrostatically tune the charge carrier population within the semiconducting channel. Positive gate voltages increase indeed the on-current, and negative gate voltages drive the device into the off-state. More clearly this can be seen in the transfer characteristic of the device in Figure 4b. At gate voltages below -10 V the device is shut off and at voltages over +0 V the device current levels off in the on-state. The hysteresis is most probably due to trapped charges in the dielectric oxide layer between the semiconducting channel and the back gate. The transfer characteristic proves that the devices are n-doped, as might be expected because of the excess of In in the InP part of the nanoneedle. The measurements show that the In/InP nanoneedles can not only be used as ready-made Schottky diodes but can even act as Schottky transistors.

\subsection*{EXPERIMENTAL DETAILS}
The InP nanoneedles were synthesized in a simple one-pot reaction, which is based on the injection of the In source (InMe3) into a hot solution containing a phosphorus source (TOP) and a stabilizing agent (TOPO). In a typical synthesis we used 10 mL (20 mmol) of trioctylphosphine (Fluka) and 1 g of TOPO (Merck) in a 25 mL three-neck flask. The solution was heated to 120$^{\circ{}}$C in vacuum to remove all volatile substances. Then, the solution was heated to an appropriate temperature under inert gas atmosphere. After reaching the reaction temperature 0.25 mmol trimethylindium
(InMe3) (Epichem) in 1 mL of octadecene (ODE) (Fluka) was quickly injected. Subsequently, the solution was allowed to cool down to room temperature. The nanowires were precipitated in a centrifuge at 4500 rpm and the sediment was washed with toluene in an ultrasonic bath for 10 min and centrifuged
again. This procedure was repeated five times. 

For varying the indium concentration we decreased the amount of trimethyl-indium while keeping TOP constant. TOPO was distilled in vacuum before use, whereas all other chemicals were used as received. 

For structural characterization the nanoneedles were investigated by XRD (Philips XfPert), high-resolution transmission electron microscopy (HRTEM, Philips CM-300 microscope, operated at 200 kV) and EDX. The samples for XRD measurements were prepared by dropping a chloroform solution of the particles on a Si wafer and evaporating the solvent. Samples for TEM analysis were prepared by drop casting a toluene solution of the particles
on a carbon-coated copper grid.

\end{document}